\documentclass{article}
\usepackage{graphicx}  
\usepackage{amsmath}   
\usepackage[compress]{cite}
\usepackage{amssymb}   
\usepackage{bm} 
\usepackage{dcolumn}
\usepackage{color}
\usepackage{mathrsfs}
\usepackage{amsfonts}
\usepackage{varioref}
\usepackage{color}
\usepackage[margin=1in]{geometry}
\RequirePackage[colorlinks,citecolor=blue,urlcolor=magenta,linkcolor=blue]{hyperref}

\def\EH{Einstein-Hilbert }
\def\LL{Lanczos-Lovelock }

\def\gr{general relativity}

\labelformat{section}{Section #1} 
\labelformat{subsection}{Section #1} 
\labelformat{subsubsection}{Section #1}
\labelformat{subsubsubsection}{Section #1}
\labelformat{equation}{Eq.~(#1)} \labelformat{figure}{Fig.~#1} 
\labelformat{subfigure}{Fig.~\thefigure#1} 
\labelformat{table}{Tab.~#1} 
\labelformat{appendix}{Appendix #1}
\title{On the physical process first law for dynamical black holes}
\author{Akash Mishra\footnote{akash.mishra@iitgn.ac.in}$~^{1}$, Sumanta Chakraborty \footnote{sumantac.physics@gmail.com}$~^{2}$, Avirup Ghosh\footnote{avirup.ghosh@iitgn.ac.in}$~^{1}$ and Sudipta Sarkar\footnote{sudiptas@iitgn.ac.in}$~^{1}$\\
\small{$~^{1}$ Indian Institute of Technology, Gandhinagar-382355, Gujarat, India}\\
\small{$~^{2}$ Department of Theoretical Physics, Indian Association for the Cultivation of Science, Kolkata-700032, India}\\
}
\date{ }  
\begin{document}
  
\maketitle
\begin{abstract}

Physical process version of the first law of black hole mechanics relates the change in entropy of a perturbed Killing horizon, between two asymptotic cross sections, to the matter flow into the horizon. Here, we study the mathematical structure of the physical process first law for a general diffeomorphism invariant theory of gravity. We analyze the effect of ambiguities in the Wald's definition of entropy on the physical process first law. We show that for linearized perturbations, the integrated version of the physical process law, which determines the change of entropy between two asymptotic cross-sections, is independent of these ambiguities. In case of entropy change between two intermediate cross sections of the horizon, we show that it inherits additional contributions, which coincide with the membrane energy associated with the horizon fluid. Using this interpretation, we write down a physical process first law for entropy change between two arbitrary non-stationary cross sections of the horizon for both general relativity and Lanczos-Lovelock gravity. 

\end{abstract}
\section{Introduction}

Black holes are inarguably the most fascinating predictions of general relativity. A remarkable resemblance exists between the laws of thermodynamics and black hole mechanics \cite{Bardeen:1973gs}.  This gravity-thermodynamics correspondence becomes non-trivial when one associates a ``temperature''  \cite{Hawking:1971vc,Hawking:1974sw} as well as an ``entropy'' with the event horizon \cite{Bekenstein:1972tm,Bekenstein:1973ur}, which is a causal boundary, that hides the central singularity of the black hole from an outside observer. Thus, the properties of the event horizon may provide some deep insight into the nature of black holes and particularly towards a theory of quantum gravity. Also, it has been shown that one can relate these thermodynamic parameters to various geometric constructions not only for event horizons, but for any null surface acting as a causal horizon, including the Rindler horizon \cite{Jacobson:1999mi, Jacobson:2003wv} (see also
 \cite{Padmanabhan:2009vy,Chakraborty:2015hna,Chakraborty:2016dwb}).
\begin{figure}[h!]
\begin{center}
\includegraphics[scale=0.6]{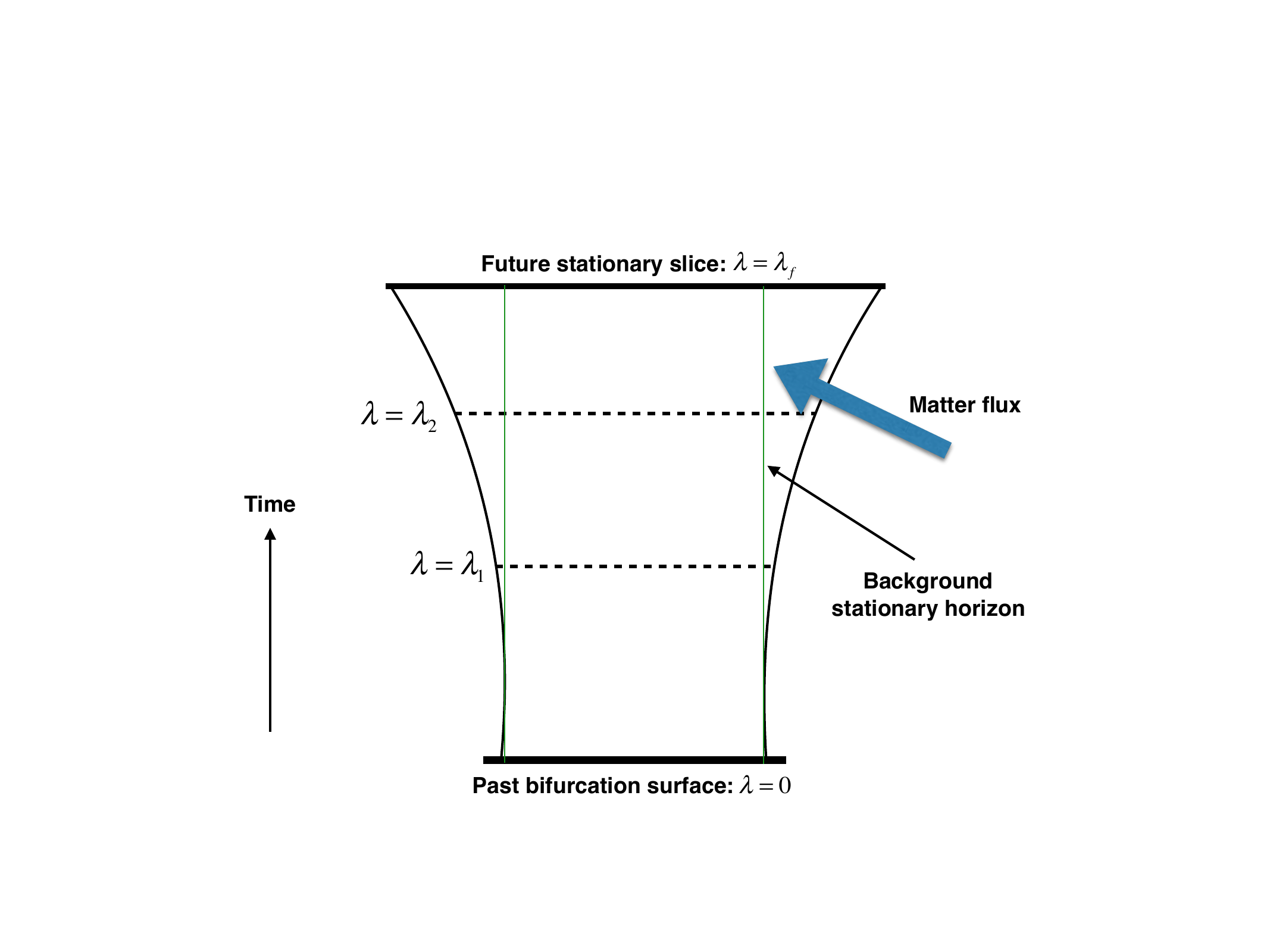}
\caption{A pictorial depiction of the geometry considered in the Physical process  version of the first law. The green line depicts the evolution of the unperturbed stationary event horizon while the black curve denotes the evolution of the perturbed dynamical event horizon. The change in area is calculated between the two slices $\lambda=0$ (the bifurcation surface) and $\lambda=\lambda_f$ (a stationary final slice) along the black curve. We would like to generalize this and calculate the change between two arbitrary slices $\lambda=\lambda_1$ and $\lambda=\lambda_2$. }
\label{fig.PP}
\end{center}
\end{figure}
The equilibrium version of the first law of black hole mechanics relates the variation of the ADM charges at asymptotic infinity to the change of the horizon area for two stationary black holes which are infinitesimally close in the phase space of solutions. Strictly speaking, this stationary state version of the first law is only applicable to a stationary event horizon with a regular bifurcation surface\footnote{ The requirement that a regular bifurcation surface be present can however be avoided in some quasi-local definitions of horizons viz. Isolated horizons\cite{Ashtekar:2001is,Ashtekar:2004cn}.}. On the other hand, another version of the first law involves the direct computation of the horizon area change when a flux of matter perturbs the horizon \cite{Hawking:1972hy} (see \ref{fig.PP}). This came to be known as the physical process first law \cite{wald1994quantum} (henceforth referred to as PPFL). Unlike the stationary state version, PPFL is local in nature and does not require the information about the asymptotic structure of the space time and is therefore expected to hold for a wide class of horizons which does not rely on the asymptotic structure. Consequently, after some initial debate regarding the applicability of PPFL in the context of Rindler \cite{Jacobson:2003wv}, it was later demonstrated, following  \cite{Amsel:2007mh,Bhattacharjee:2014eea}, that the physical process version of first law indeed holds for Rindler horizon, or for that matter, any bifurcate Killing horizon. 

Consider a situation, in which a black hole is perturbed matter influx with stress energy tensor $T_{ab}$ and the black hole finally settles down to a new stationary horizon in the future. Then the PPFL which determines the change of the horizon area $A_{\rm H}$, is given by,
\begin{align}\label{BH_Dyn_01}
\frac{\kappa}{2\pi}\Delta \left(\frac{A_{\rm H}}{4}\right)=\int_{{\cal H}}  T_{ab}\,\xi^{a}\, d\Sigma^b~.
\end{align}
Here $d\Sigma^b = k^{b} \,dA\,d\lambda$ is the surface area element of the event horizon and $k^a = \left( \partial / \partial \lambda \right)^a$ stands for the null generator of the horizon. The integration is over the dynamical event horizon and the affine parameter $\lambda$ varies from the bifurcation surface (set at $\lambda = 0$) to the future stationary cross section at $\lambda = \lambda_f$. Also, the background event horizon is a Killing horizon with the Killing field $\xi^a$ being null on the background horizon. On the background horizon surface, it is related with the affinely parametrized horizon generator ($k^a$) as $\xi^a = \lambda \kappa k^a$, where $\kappa$ is the surface gravity of the background Killing horizon. It is important to note that the derivation of the above result crucially hinges to the fact that the terms quadratic in expansion and shear of the null generator $k^{a}$ can be neglected, since the process has been assumed to be sufficiently close to stationarity. This approximation ensures that there will be no caustic formation in the range of integration.

For the generalization of these results to higher curvature theories of gravity, first one needs to define a suitable notion of entropy of the horizon. For stationary event horizon with a regular bifurcation surface, the entropy of black hole is given by Wald's entropy formula and the stationary state version of first law directly follows\cite{Wald:1993nt,Iyer:1994ys}. However, when applied to a non-stationary cross section, the Wald's entropy turns out to be ambiguous\cite{Jacobson:1993vj} and as far as the PPFL is concerned, it no longer can be used as a good measure of entropy. Hence, it is very important to understand, how the ambiguities affect the PPFL and consequently the entropy increase theorem at various orders of perturbation. In this article we prove that, the ambiguities in Wald entropy doesn't affect the PPFL at first order perturbation, when integrated from a bifurcation surface to a stationary surface. Therefore if the horizon is involved in a quasi stationary process such that the perturbation remains small throughout the evolution, then an unambiguous version of PPFL can be obtained by the variation of Wald entropy, independent of the ambiguities in the Noether charge construction.
However, the second order variation of entropy turns out to be affected by ambiguities.  For these general theories of gravity the above ambiguities can however be fixed such that linearized second law holds, leading to modified expressions for the horizon entropy \cite{Rogatko:2002eu,Chatterjee:2011wj,Kolekar:2012tq}.

Another interesting aspect transpires when one tries to link it with the horizon membrane, where the coefficient of $\theta^2$ and $\sigma^2$ in the Raychaudhuri equation can be identified as bulk and shear viscosity respectively of the fictitious fluid living on the horizon. Therefore by a careful study of the PPFL in a more general form, i.e., by keeping terms quadratic in expansion and shear, one can comment on the behaviour of the membrane fluid in a more general context \cite{Fairoos:2018pee}.

The physical process first law therefore relates the total change of entropy from the  bifurcation surface to a final slice due to the matter flux. If we assume that the black hole horizon is stable under perturbation, then the future slice can always be taken to be stationary with vanishing expansion and shear and the initial cross section can be set at the bifurcation surface ($\lambda=0$). In fact, the choice of these initial and final states are necessary for this derivation of the physical process first law, to make some boundary terms vanish. Here we will also provide a derivation of the physical process version of the first law between two arbitrary non equilibrium cross sections of the dynamical event horizon and properly interpret the additional boundary terms appearing in \ref{BH_Dyn_01} for entropy change. We will show how these boundary terms are related to the energy of the horizon membrane arising in the context of the black hole membrane paradigm. Moreover, we will establish that such a correspondence transcends beyond \gr\ and also holds true for \LL theories of gravity. 

The paper is organized as follows: In the first section we start by describing the horizon geometry and set up the notations and conventions. Then, a general expression for the PPFL is illustrated and possible limiting cases are also discussed. At first order we obtain an equation for the flow of entropy. We start \ref{GSPPFL} with a very brief discussion on ambiguities in the Wald's entropy formula, when evaluated on a non-stationary cross-section of the horizon. In \ref{BEA} we show that, at first order of perturbation the ambiguities do not affect the PPFL obtained from Wald entropy. In subsequent sections we will generalize PPFL to dynamical situations (i.e., for arbitrary cross-sections of the horizon) using the stress tensor of the horizon fluid in membrane paradigm. The dynamical version of the physical process first law for Einstein-Gauss-Bonnet theories of gravity will be derived in \ref{Sec_03} and finally the generalization of the same to general \LL theories of gravity will be performed in \ref{Sec_04}. We conclude with a discussion on our results as well as with some future outlook.  
\section{General Structure of PPFL}\label{GSPPFL}

We start this section by describing the geometry of the event horizon. The event horizon $H$ of a stationary black hole in $D$ spacetime dimensions is a null hypersurface generated by a null vector field $k^a=(\partial / \partial\lambda)^a$ with $\lambda$ being an affine parameter. The cross section ($\mathcal{H}$) of the event horizon, which is a co-dimension two spacelike surface, is given by each $\lambda = constant$ slice. Being a co-dimension two surface, $\mathcal{H}$ posses two normal direction. One of them is the null normal $k^{a}$ and the other corresponds to an auxiliary null vector $l^{a}$ defined on $\mathcal{H}$ such that $k_a l^a =-1$. Then, the induced metric on the horizon cross section takes the form, $h_{ab} = g_{ab}+k_a l_b + k_b l_a$. Taking $x^A$ to be the coordinates on $\mathcal{H}$, $(\lambda,x^A)$ spans the horizon. The event horizon of a stationary black hole is also a killing horizon, which is defined as the set of points on which the  killing vector $\xi^a$ is proportional to the horizon generator $k^a$, i.e., $\xi^a = \lambda \kappa k^a$. We consider the background event horizon, on which the perturbation is introduced, to be stationary with respect to the killing vector $\xi^a$.
Here $\kappa$ represents the surface gravity of the horizon and defined as $\xi^a\nabla_a \xi^b \stackrel{\mathcal H}{=} \kappa \xi^b$.

\begin{figure}
\centering
\includegraphics[scale=0.6]{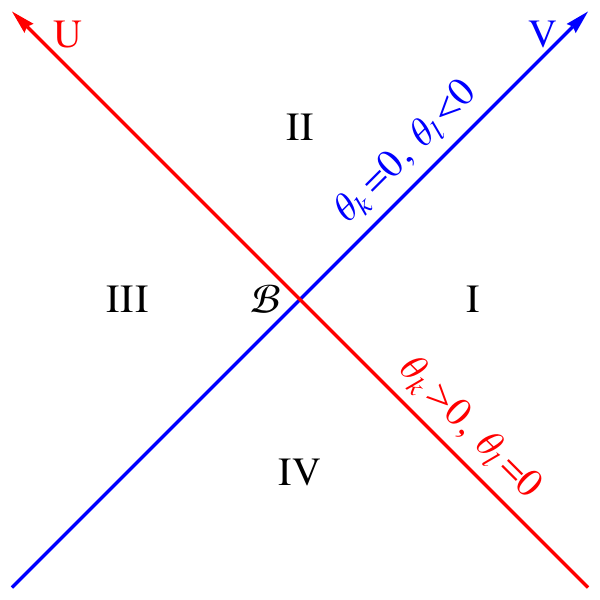}
\caption{The point $\mathcal{B}~(U=0,V=0)$, a $(D-2)$-dimensional cross-section of the Horizon represents the Bifurcation surface, where $\theta_k,\theta_l=0$.}\label{Kruskal}
\end{figure}

With this setting, we define the expansion and shear of the horizon to be the trace and traceless symmetric part of the extrinsic curvature and denoted as $(\theta_k,\sigma^k_{ab})$ and $(\theta_l,\sigma^l_{ab})$ with respect to $k^a$ and $l^a$ respectively. Taking $h$ to be the determinant of the induced metric $h_{ab}$, the expansion $\theta_k$ of horizon can be written as,
\begin{equation}
\theta_k = \frac{1}{\sqrt{h}} \frac{d}{d\lambda}\sqrt{h}
\end{equation}
Then, the evolution of $\theta_k$ along the horizon with respect to the affine parameter $\lambda$ is governed by the Raychaudhuri equation,
\begin{equation}
\frac{d\theta_k}{d\lambda} = -\frac{1}{D-2}\theta_k^2 -\sigma^k_{ab}\sigma^{ab}_k - R_{ab}k^a k^b
\end{equation} 
An important notion that will play a significant role throughout our discussion is a bifurcation surface. A bifurcation surface is a $(D-2)$ dimensional spacelike surface $\mathcal{B}$, on which the killing field $\xi^a$ identically vanishes. Also $\mathcal{B}$ is the surface on which the past and future horizon intersects. For our purpose it is convenient to choose $\mathcal{B}$ to be at $\lambda=0$, otherwise it is completely arbitrary. The traditional illustration of the PPFL and the horizon entropy increase theorem extensively rely on the existence of a bifurcation surface. But, in general the bifurcation surface is not a part of black hole space time formed by the gravitational collapse of an object. However, if the geodesics that generate the horizon are complete to the past, one can always have a bifurcation surface at some earlier $\lambda$. This can be realized by the maximal extension of the black hole space-time. For instance, no notion of bifurcation surface exist in the Schwarzschild space-time. Nevertheless, in its maximal extension i.e., in the Kruskal space-time, the 2-sphere at $U=0,V=0$ represents a bifurcation surface, as indicated in \ref{Kruskal} . A simple calculation leads to the following expressions of the expansion coefficients along $k$ and $l$,
\begin{equation}
\theta_k \propto U;\qquad \theta_l \propto V 
\end{equation}
Hence, at the future horizon $U = 0$, the expansion $\theta_k = 0$ and at the past horizon where $V = 0$, we have $\theta_l = 0$. But, at the bifurcation surface $(U=0,V=0)$ both $\theta_k$ and $\theta_l$ vanishes. Also the shear can be shown to be vanishing on $\mathcal{B}$. As a result, $\theta_k$ and $\sigma_k$ are of first order in perturbation, i.e., $\mathcal{O}(\epsilon)$, while $\theta_l$ and $\sigma_l$ are of zeroth order, with $\epsilon$ referring to the strength of perturbation everywhere on the future event horizon. However, since $\theta_l$ and $\sigma_l$ vanishes at the bifurcation surface of the stationary black hole, they must be of at least $\mathcal{O}(\epsilon)$ only at $\mathcal{B}$. This result is indeed a property of the bifurcation surface itself and independent of the physical theory one considers. Hence it also generalizes beyond general relativity and holds for higher curvature theories as well. In summary we have,
$\theta_k,\, \sigma_k,\, R_{ab}k^{a}k^{b},\, d\rho/d\lambda\backsim \mathcal{O}(\epsilon)$ and $\theta_l,\sigma_l \backsim \mathcal{O}(\epsilon)$ at $\mathcal{B}$. As a result, terms like $\theta_k \theta_l \backsim \mathcal{O}(\epsilon^2)$ only at the bifurcation surface of the background stationary horizon.

Having defined the geometry of the horizon, we are now set to illustrate the physical process version of first law for an arbitrary diffeomorphism invariant theory of gravity in its most general form. In order to discuss the PPFL, first one needs to define some suitable notion of entropy on the horizon. But, since we consider theories beyond general relativity here, the standard formula for entropy, as being proportional to area, no longer holds. Nevertheless, whatever the expression for entropy might be, it must be some local functional density integrated over the horizon. Hence we start by considering the following expression for horizon entropy:
\begin{equation}
S = \frac{1}{4} \int_{\mathcal{H}} (1+\rho) \sqrt{h}\;d^{D-2}x ,\label{BH entropy}
\end{equation}
where $\rho$ is some entropy density constructed locally on the horizon and contains the higher curvature contributions. The area-entropy relation in general relativity limit can be obtained by setting $\rho=0$. The field equation in such a general theory can always be written as,
\begin{equation}
R(k,k)+H(k,k) = 8\pi T(k,k)\label{general equation}
\end{equation}
where, $R(k,k)\equiv R_{ab}k^{a}k^{b}$ and $T(k,k)=T_{ab}k^{a}k^{b}$ along with $H(k,k)\equiv H_{ab} k^a k^b$ represents the deviation from general relativity. With this setting, let us now compute the variation of the entropy along the horizon generator $k^{a}$, in response to some influx of matter,
\begin{align}
\Delta S(\rho) = \frac{1}{4}\int_{\mathcal{H}} d^{D-2}x \int \frac{d}{d\lambda}[(1+\rho)\sqrt{h}]\;d\lambda
\end{align}
This can be further simplified by integrating by parts and finally the change in entropy takes the form,
\begin{align}
\Delta S(\rho) =&\frac{1}{4}\left(\int  dA\; \lambda \Theta_{k}\right)_{\lambda_1}^{\lambda_2} + 2\pi \int dA \;d\lambda\; \lambda T(k,k) + \frac{1}{4}\int dA\; d\lambda\; \lambda\left[-\left(\frac{D-3}{D-2}\right)(1+\rho)\theta_k^2 +(1+\rho)\sigma^2 \right]\nonumber \nonumber\\&-
\frac{1}{4} \int dA\;d\lambda\; \lambda \left(\frac{d^2\rho}{d\lambda^2} + 2\theta_k \frac{d\rho}{d\lambda} -\rho R(k,k) + H(k,k)\right),\label{general_PPFL}
\end{align}
where $\Theta_{k}= \theta_k + \rho \theta_k + \frac{d\rho}{d\lambda}$ and $T(k,k)$ is understood as the $(k,k)$ component of the matter stress tensor. 

We would like to emphasize that \ref{general_PPFL} represents the most general form of the variation of entropy along the null generator and no assumption regarding the strength of the perturbation or the range of integration has been made throughout the derivation. With this general expression in hand, one can now discuss the change in entropy at various order of perturbation. Since terms like $\theta_k^2,\,\sigma_k^2$ and $\theta_k (d\rho/d\lambda)$ are of $\mathcal{O}(\epsilon^2)$, they do not contribute to the first order variation. Hence, truncating the general result upto first order in perturbation, we find that the change in entropy takes the form,
\begin{equation}
\Delta S^{(1)}(\rho) =\frac{1}{4}\left(\int dA\; \lambda \Theta_{k}\right)_{\lambda_1}^{\lambda_2} + 2\pi \int dA \;d\lambda\; \lambda T(k,k) -
\frac{1}{4} \int dA\;d\lambda\; \lambda \left(\frac{d^2\rho}{d\lambda^2}  -\rho R(k,k) + H(k,k)\right)\label{first o}
\end{equation}
For further simplification lets evaluate the last integrand of the above equation for some simplified models, say $f(R)$ theory, for which, the field equation takes the form,
\begin{equation}
f'(R)R_{\mu\nu}-\frac{1}{2}g_{\mu\nu}f(R)+g_{\mu\nu}\square f'(R) -\triangledown_\mu \triangledown_\nu f'(R) = 8\pi T_{\mu\nu}
\end{equation}
Note that, for $f(R)$ theory, $f'(R)$ represents the modifications to the entropy density over and above the \EH expression, i.e., $\rho=f'(R)-1$, where the prime denotes first derivative w.r.t the Ricci scalar $R$. Also, one can always rewrite the field equation for $f(R)$ theory in the form of \ref{general equation}, with 
\begin{align}
H(k,k)= f'(R)R(k,k) - k^a k^b \nabla_a\nabla_b f'(R)
\end{align}
Substitution of the above expression for $H(k,k)$ results into the following identity for $f(R)$ theories,
\begin{equation}
\frac{d^2\rho}{d\lambda^2}  -\rho R(k,k) + H(k,k)=0\label{entropy flow}
\end{equation}
If \ref{entropy flow} is a property of the entropy density itself, then our expression for the change in entropy upto first order as given in \ref{first o} will take a very simplified form. At this stage it seems a mere coincident, but as will be pointed out later, one can show the above equation to hold for arbitrary order Lovelock theory as well. Motivated by these results, we argue that, \ref{entropy flow} is a general property of the entropy density and holds in an arbitrary diffeomorphism invariant theory of gravity. Note that, the above equation for the flow of entropy density is a result that holds only up to first order in perturbation, i.e., it may be proportional to terms that are at least quadratic in expansion and shear. Hence in general \ref{entropy flow} is indeed of the form,
\begin{equation}
\frac{d^2\rho}{d\lambda^2}  -\rho R(k,k) + H(k,k)= O(\epsilon^2)
\end{equation}
With this result, the first order variation of the entropy simplifies to,
\begin{equation}
\Delta S^{(1)}(\rho) =\frac{1}{4}\left(\int dA\; \lambda ~\Theta_{k}\right)_{\lambda_1}^{\lambda_2} + 2\pi \int dA \;d\lambda\; \lambda\, T(k,k)\label{firstorder-entropy}
\end{equation}
In the general relativity limit ($\rho\rightarrow 0$) $\Theta _{k}\rightarrow \theta_k$ and hence the boundary term doesn't contribute when integrated from a bifurcation surface to a stationary slice. 
This is because we have set the bifurcation surface to be at $\lambda=0$ and the final stationary cross-section has vanishing expansion. 
But, when the PPFL is evaluated between two non equilibrium slice $\lambda_1$ and $\lambda_2$, the boundary term $\theta_k$ turns out to be non vanishing. We will discuss the issue of arbitrary cross sections later and at this stage it will be worthwhile to briefly mention the standard derivation of the physical process first law, in which the entropy change is computed using two assumptions:
\begin{itemize}
\item The horizon possesses a regular bifurcation surface in the asymptotic past, which is set at $\lambda = 0$ in our coordinate system.

\item The horizon is stable under perturbation and eventually settle down to a new stationary black hole. So, all Lie derivatives with respect to horizon generators vanish in the asymptotic future.
\end{itemize}
The first assumption is technical in nature. Note that, in a realistic scenario of black hole formation by gravitational collapse, the bifurcation surface is not the part of the physical space time. Therefore, it is desirable to have a formulation of the physical process first law without this assumption.

The second assumption is motivated by the cosmic censorship conjecture which asserts that the black hole horizon must be stable under perturbation, so that expansion and shear vanish in the asymptotic future at $\lambda = \lambda_f$. This is in principle similar to the assertion that a thermodynamic system with dissipation ultimately reaches an equilibrium state. This is a desirable property of the black hole horizon. Moreover, while deriving the physical process first law, we are already neglecting higher order terms and this requires that small perturbations remain small throughout the region of interest.  This is equivalent to not having any caustic formation on any portion of the dynamical horizon. Under these assumptions, the boundary term in (\ref{firstorder-entropy}) drops out, provided the generalized expansion $\Theta_k$ goes to zero in the future faster than the time scale $1/\lambda$, and we are left with,
\begin{align}\label{BH_Dyn_03}
\Delta S&= 2 \pi \int _{\lambda = 0}^{\lambda _f}\lambda ~d\lambda\, d^{2}x\,\sqrt{q}\,T_{ab}k^{a}k^{b}~.
\end{align}
Subsequently identifying the background Killing field as $\xi^a = \lambda\, \kappa \,k^a$ one can rewrite the above equation as,
\begin{align}
\frac{\kappa}{2\pi}\Delta S=\int_{{\cal H}}  T_{ab}\,\xi^{a}\, d\Sigma^b~.
\end{align}
This completes the standard derivation of what is known as the integrated version of the physical process first law described by \ref{BH_Dyn_01}. If the matter field satisfies the null energy condition, then one will have $T_{ab}k^{a}k^{b}\geq 0$. As a consequence, it will follow that the total change in entropy between the boundary slices is also positive semi-definite. In comparison with the stationary version of the first law, the PPFL is local and independent of the asymptotic structure of the space time. In fact, the relationship between these two versions are not totally straightforward. In the next section, we would like to understand how these two approaches are related to each other.
\section{Stationary comparison version and Physical process Law}

The version of the first law for black holes which deals with the variations in the phase space of the theory is called the stationary comparison version of the first law. This version is originally derived in \cite{Bardeen:1973gs} and then generalized to arbitrary diffeomorphism invariant theories by Wald and others \cite{Wald:1993nt,Iyer:1994ys}. This version starts with a stationary black hole solution. In general relativity, the strong rigidity theorem implies that in a stationary spacetime the event horizon is also a Killing horizon. Although there is a generalization of this theorem beyond general relativity, it is always assumed that the black hole event horizon is also a Killing horizon, such that there is a timelike Killing vector field outside, which becomes null on the horizon. It is also assumed that the horizon possesses a regular bifurcation surface in the past. This assumption is necessary for Wald's construction. Also, the existence of the past bifurcation surfaces lead to a constant surface gravity independent of the theory of gravity \cite{Racz:1995nh}. Then, the stationary comparison version compares two such nearby stationary solutions in the phase space which differ infinitesimally in ADM mass and relates the change of ADM mass $\delta M$ to the entropy variation $\delta S$ as,
\begin{equation}
\frac{\kappa}{2 \pi} \delta S = \delta M
\end{equation}
The variation $\delta$ is to be understood in the space of solutions. 

To understand the relationship with PPFL, consider a time dependent black hole solution: say for simplicity, a spherically symmetric Vaidya black hole which is accreting radiation. The metric for such a space time is \cite{Wald:1984rg},
\begin{equation}
ds^2 = - \left(1 - \frac{ 2 M(v) }{ r} \right)dt^2 + 2 dv \,dr + r^2 d\Omega^2.
\end{equation}
The Vaidya space time is an excellent scenario to study the physical process first law. The area of the event horizon is increasing due to the flux of the ingoing matter. The rate of change of the time dependent mass $M(v)$ represents the energy entering into the horizon. But, although $M(v)$ is changing with time, the ADM mass of the spacetime is constant, evaluated at the spacelike infinity: $M_{\textrm{ADM}} = M(v \to \infty)$. In fact, by definition, there is no physical process which can change the ADM mass of a space time. Therefore, the relationship between the PPFL and the stationary comparison version is somewhat subtle. 

To understand the relationship, we consider the Vaidya space time, as a perturbation over a stationary black hole of ADM mass $m$. Therefore, we assume $M(v) = m + \epsilon \, f(v)$. The parameter $\epsilon$ signifies the smallness of the perturbation. Note that, the background spacetime with ADM mass $m$ is used only as a reference, it does not have any physical meaning beyond this. In the absence of the perturbation, the final ADM mass would be same as $m$. Therefore, we may consider the process as a transition from a black hole of ADM mass $m$ to another with ADM mass $M_{\textrm{ADM}}$ and this allows us to relate the PPFL to the stationary comparison law.

In case of ordinary thermodynamic system, the entropy is a state function and its change is independent of the path. Therefore, we can calculate the change of entropy due to some non equilibrium irreversible process between two equilibrium states by using a completely different reversible path in phase space. In black hole mechanics, the stationary comparison law can be thought as the change of entropy along a reversible path in the space of solutions, whereas the PPFL is a direct irreversible process. The equality of the entropy change for both these process therefore shows that the black hole entropy is indeed behaving as that of true thermodynamic entropy. 

Having understood the relationship between these two versions of the first law for black holes, we will now study the ambiguities of Wald's construction for black hole entropy and how PPFL is affected by such ambiguities.
\section{Black hole Entropy and Ambiguities}\label{BEA}

The entropy of stationary black holes with a regular bifurcation surface in an arbitrary diffeomorphism invariant theory of gravity  
is given by Wald's formula\cite{Wald:1993nt} as,
\begin{equation}
S_{W} = -2\pi\int_{\mathcal{B}} \frac{\partial L}{\partial R_{abcd}}\epsilon_{ab}\epsilon_{cd} \,\sqrt{h} \,d^{D-2}x = \int_{\mathcal{B}} (1+\rho_w) \sqrt{h} \,d^{D-2}x\label{Wald entropy}
\end{equation}
Where $\epsilon_{ab} = k_a l_b -k_b l_a$ is the bi-normal to the bifurcation surface. Note that, the construction of Wald entropy formula crucially depends on the existence of a bifurcation surface. However, as pointed out by\cite{Jacobson:1993vj}, the Wald entropy remains unaffected even when evaluated on an arbitrary cross section of a stationary event horizon. This can be realized from the fact that, in a stationary black hole space time, every cross section of the horizon is isomorphic to other. Since, in the Wald's construction, the Noether charge $(D-2)-$ form is constructed locally on the $(D-2)$ dimensional horizon cross section, one would expect the Noether charge entropy to be the same, irrespective of which cross section it is evaluated on. 

Also, as discussed in\cite{Iyer:1994ys,Jacobson:1993vj}, the ambiguities in the Noether charge construction doesn't affect the Wald entropy in case of stationary black hole. However, if the horizon is involved in a dynamical process i.e., for non stationary black holes, the Wald entropy formula no longer holds and turns out to be ambiguous upto addition of terms of the form,
\begin{equation}
\Delta S_w = \int \Omega~dA\; ,
\end{equation}
where $\Omega =(p \theta_k \theta_l + q \sigma_k\sigma_l)$ and $\sigma_k \sigma_l = \sigma_{ab}^{k}\sigma^{ab}_l$. 
Note that, terms in $\Omega$ contains equal number of $k$ and $l$ indices and hence combine to produce a boost invariant object, although they individually transform non-trivially under boost. The coefficients $p$ and $q$ can be fixed by demanding that the entropy be locally increasing\cite{Bhattacharjee:2015yaa, Bhattacharjee:2015qaa}.

Comparing \ref{BH entropy} and \ref{Wald entropy} and taking into account the ambiguities, we can identify $\rho = \rho_w + \Omega$. This identification essentially means that, the black hole entropy for a non-stationary horizon slice can always be expressed as the entropy of the stationary horizon plus ambiguities. Note that the black hole entropy coincide with Wald entropy in the stationary limit. Now, we would like to ask a definite question: how does the physical process law get affected by the ambiguities in the Noether charge construction? We will show that as in the case of the stationary version, the physical process law for linear perturbations is also independent of these ambiguities, provided we consider the entropy change from the past bifurcation surface to the final stationary cross section. To see this, consider the first order variation of the  Wald entropy,
\begin{equation}
\Delta S^{(1)}(\rho_w) =\frac{1}{4}\left(\int dA\; \lambda~ \Theta_{k}{(\rho_w)}\right)_{\lambda_1}^{\lambda_2} + 2\pi \int dA \;d\lambda\; \lambda T(k,k)  -
\frac{1}{4} \int dA\;d\lambda\; \lambda \left(\frac{d^2\rho_w}{d\lambda^2}  -\rho_w R(k,k) + H(k,k)\right)\label{first order Wald}
\end{equation}
Here we emphasize that, the equation for the flow of entropy that is obeyed by $\rho$ may not hold for $\rho_w$ because of the presence of ambiguity. Now, we are interested in the difference between the change in black hole entropy to the change in Wald entropy upto first order in expansion and shear. A straightforward calculation shows,
\begin{equation}
\Delta S^{(1)}(\rho) -\Delta S^{(1)}(\rho_w) = \frac{1}{4}\int dA \lambda \left( \frac{d\Omega}{d\lambda} +\Omega\theta_k \; \right)\Bigg|_{\lambda_1}^{\lambda_2}- \frac{1}{4} \int dA\, d\lambda \left(\lambda \frac{d^2\Omega}{d\lambda^2}\right)
\end{equation}
Where we have neglected the term $\Omega R(k,k)$, which is of $\mathcal{O}(\epsilon^2)$. Also $\Omega\theta_k$ is of $\mathcal{O}(\epsilon^2)$ and does not contribute to the first order variation. Simplifying it further one can obtain,
\begin{equation}
\Delta S^{(1)}(\rho) -\Delta S^{(1)}(\rho_w) = \frac{1}{4}\int dA\; \Omega \Big|_{\lambda_1}^{\lambda_2}\label{entropy difference}
\end{equation}
The above equation represents the difference in the change in black hole entropy and Wald entropy as a boundary term evaluated between two arbitrary non equilibrium horizon slices at $\lambda_1$ and $\lambda_2$. Since $\Omega$ involves shear and expansion, the right hand side of \ref{entropy difference} doesn't vanish in general, when evaluated on non-stationary cross sections. However, consider the case when the background stationary black hole has a regular bifurcation surface and the lower limit of the integration is at that surface set at $\lambda = 0$. Then, as discussed in the previous section, terms like $\theta_k \theta_l$ are of second order in perturbation and therefore $\Omega$ turns out to be $\mathcal{O}(\epsilon^2)$ and does not contribute to the linear order calculation. Hence, up to first order, the ambiguities does not affect the PPFL when integrated from a bifurcation surface to a stationary surface. This is analogous to the case of stationary version of first law as proven in \cite{Iyer:1994ys,Jacobson:1993vj}.

This result however doesn't hold when second order perturbations are considered. Taking the difference between $\Delta S(\rho)$ and $\Delta S(\rho_w)$ by keeping terms upto $\mathcal{O}(\epsilon^2)$ one can obtain,
\begin{equation}
\Delta S^{(2)}(\rho) -\Delta S^{(2)}(\rho_w) = \frac{1}{4}\int dA\, \left(\lambda \theta_k \Omega + \Omega\right)\Big|_{\lambda_1}^{\lambda_2} + \frac{1}{4} \int dA\, d\lambda \left(\lambda \theta_k \frac{d\Delta}{d\lambda} +\lambda  \Omega R(k,k)-\Omega \theta _{k}\right)
\end{equation}
Unlike first order, the difference in the change in black hole entropy and Wald entropy is given by a boundary term and a bulk integral. As a result, any conclusion about the change of black hole entropy beyond linearized perturbation requires the resolution of these ambiguities. In fact, if we demand an instantaneous second law, such that the entropy is increasing at every cross section, to hold  beyond general relativity, we can fix the ambiguities and find the appropriate black hole entropy\cite{Bhattacharjee:2015yaa, Bhattacharjee:2015qaa}. Then, it is possible to study the higher order perturbations and obtain the transport coefficients related to the horizon \cite{Fairoos:2018pee}. \\

So far, we have only considered the integrated version of PPFL, where the entropy change is calculated between the past bifurcation surface and the future stationary regime. In the next sections, we would like to understand the PPFL for two arbitrary cross sections of the horizon. In order to do so, we would require some results from the membrane paradigm formulation for black hole. We therefore, briefly describe the membrane construction for the event horizon, in the next section.

\section{Membrane Paradigm for Black holes}

The membrane paradigm is an effective description of black hole physics from the point of view of an outside observer. In membrane paradigm, one models the black hole horizon by a membrane of a fictitious fluid living on the horizon surface. Thus the interaction of the black hole with outside matter is mimicked by the interaction of the horizon fluid with the matter and hence is related to the transport coefficients of the fluid \cite{Thorne:1986iy}. 

It is possible to derive a membrane stress tensor using a variational principle in a space time with an inner-boundary such as black hole horizon. For example, varying the Einstein-Hilbert action functional with the Gibbons-Hawking surface term \cite{Gibbons:1976ue,Parattu:2016trq} leads to boundary contributions which vanish when Dirichlet boundary condition is imposed. But, in the presence of a black hole horizon as the inner boundary, the contribution at the inner boundary makes the variational principle ill defined. As a result, to derive the field equation,  a new contribution is added at the horizon to make the on-shell action stationary under a variation. This new contribution can be interpreted as due to some fictitious matter living on the stretched horizon, a time like surface just outside the horizon \cite{Parikh:1997ma}. The matter stress tensor on the time like membrane is then shown to be of the following form \cite{Thorne:1986iy,Parikh:1997ma}:
\begin{align}
t^{ab} = p \, h^{ab} + 2 \, \eta \,\sigma^{ab} + \zeta \,\theta _{k} \,h^{ab},
\end{align}
where the constants appearing in the above stress tensor has the following expressions: $ p = \kappa$, $\eta=1/16\pi $ and $\zeta = - 1/16\pi$ for general relativity. For other theories of gravity the form of the stress energy tensor remains unchanged. However the coefficients get modified \cite{Jacobson:2011dz, Kolekar:2011gg, Zhao:2015inu}. 

This stress tensor resembles that of a viscous fluid with shear viscosity $\eta$ and a negative bulk viscosity $\zeta$. Using this stress tensor and the tangent to the stretched horizon $u^a$, it is possible to obtain the following energy density, $\Sigma \equiv t_{ab}u^{a}u^{b}=-\theta_{k} /8\pi$, where a limit from the stretched horizon to the true horizon has been taken and the result is derived in the non-affine parametrization. Note that, because of Hawking's area theorem, the expansion $\theta_{k}$ is always positive, which makes the energy of the membrane negative as long as matter obeys null energy condition. Also, this energy for the membrane vanishes for a stationary horizon.

The negativity of the membrane energy can be explained as follows \cite{Thorne:1986iy}: The event horizon exhibits a teleological property and one has to impose future boundary conditions to ensure the stability under perturbation. Consequently, the horizon tends to expand even before matter energy actually hits the horizon. Since the matter has energy which is positive, an expanding event horizon must have negative energy so that once the matter has passed through the horizon, the total energy at equilibrium becomes zero. The same teleological condition also results in the negative bulk viscosity \cite{Padmanabhan:2010rp,Kolekar:2011gw,Eling:2011ms,Bhattacharya:2015qkt}. \\

\section{Physical process first law and the membrane paradigm}\label{PPFLMP}

We seek to formulate a generalization of the physical process first law for two arbitrary dynamical cross section of the horizon and like to understand the physics behind the above process when we integrate and evaluate the entropy change between two arbitrary slices at $\lambda = \lambda_1 $ and $\lambda = \lambda_2$. Since the boundary terms already have a contribution from $\theta _{k}$, to first order the affine parameter $\lambda$ can always be taken to be the one for the stationary background. Also, we switch to non-affine parametrization of the null generators and use following relation between the expansion scalars in different parametrization,
\begin{align}\label{ES}
\theta _{k}^{\rm (affine)}=\frac{1}{\kappa \lambda}\theta~,
\end{align}
where $\kappa$ stands for the surface gravity in the background stationary black hole and hence can be treated as constant, while  $\theta$ represents the expansion in non-affine parametrization \footnote{This will be the general strategy followed in the rest of the paper. The quantity without the subscript $k$ will correspond to the quantity in non affine parametrization.}. Using the above relation between expansion scalars in \ref{ES} and multiplying both sides of the equation by the surface gravity $\kappa$, we obtain the entropy change between two arbitrary slices as, 
\begin{align}\label{BH_Dyn_04}
\frac{\kappa}{2\pi}\Delta S &=\Delta \left[\frac{1}{4}\int d^{2}x~\theta \,\sqrt{q}\right]
+ \int_{{\cal H}}  T_{ab}\,\xi^{a}\, d\Sigma^b~,
\end{align}
where all the quantities correspond to non-affine parametrization. The last term in the left hand side is identified as the energy flux $\Delta Q$ flowing into the horizon. It is now clear that the above equation indeed has a thermodynamic interpretation provided one can interpret $\theta$ to be proportional to some sort of energy. It turns out that this can indeed be done in connection with the black hole membrane paradigm. This will enable us to write down a thermodynamic physical process first law for the entropy change between two arbitrary dynamical slices of the horizon. Before delving into details of this, let us first briefly recollect the basics of membrane paradigm. 

A look at the first term in the right hand side of \ref{BH_Dyn_04} is sufficient to convince that it is the change of this membrane energy. Using the expression of the membrane energy, $\Sigma =-\theta /8\pi$, we obtain,
\begin{align}\label{BH_Dyn_05}
\frac{\kappa}{2\pi}\Delta S  =\Delta E+  \int_{{\cal H}}  T_{ab}\,\xi^{a}\, d\Sigma^b~;
\qquad
E=-\int d^{2}x~\sqrt{q}\,\Sigma~,
\end{align}
which is the first law in a dynamical context, where unlike \ref{BH_Dyn_01}, we have an additional energy term $E$, which has its origin in the energy of the horizon fluid arising from membrane paradigm. Since the membrane energy density $\Sigma$ is negative it immediately follows that the energy $E$ defined here is positive. In the context of integrated version where the range of integration is from the bifurcation surface to future stationary slice,  it follows that the membrane energy identically vanishes at both the limits and one immediately arrives at the original physical process version of the first law. 

If we identify $\kappa / 2 \pi$ as the Hawking temperature $T_H$ of the background stationary horizon, we have the following from of the physical process first law: $T_H  \Delta S = \Delta E + \Delta Q$ for changes between two arbitrary non stationary cross sections of the dynamical event horizon. Therefore, we have explicitly shown that one can indeed write down a well defined physical process version of the first law for any two arbitrary cross sections of the horizon and the energy density of the horizon fluid in the membrane paradigm plays a crucial role in the analysis. This result illuminates a connection between the laws of black hole mechanics and membrane paradigm in a dynamical context. 

So far, we have only considered the first law for \gr. As a result, our horizon entropy is taken to be a quarter of the area. But, the integrated version of the physical process law can be generalized beyond Einstein gravity, in particular to higher curvature theories of gravity as well. For example, it is possible to write down a physical process first law for \LL class of theories \cite{Kolekar:2012tq} where the entropy is no longer proportional to the area, rather is a complicated function of the horizon geometry \cite{Jacobson:1993xs}. As a result, the extra terms arising in the expression of entropy change, between two arbitrary cross sections, is different. We would like to understand if our interpretation of the boundary terms as the energy associated with horizon membrane holds true even in such a situation. We answer this in the affirmative in the next sections, where we show how the integrated physical process first law for \LL gravity can also be generalized for entropy change between arbitrary slices. In this case also, the boundary terms can be interpreted as the energy associated with the corresponding membrane at the horizon. This result shows that our generalization transcends beyond \gr\ and have a much broader applicability.

In the next section, we will consider the case of Einstein-Gauss Bonnet gravity, the first non-trivial \LL term. We will depict how the physical process first law for arbitrary cross sections can be obtained for Einstein-Gauss-Bonnet gravity and shall show its relationship with the corresponding membrane energy. We will extend our analysis to the full \LL gravity in later sections.
\section{First law for dynamical black holes in Einstein-Gauss-Bonnet gravity}\label{Sec_03}

We have derived the first law for dynamical black holes in a quasi-stationary scenario albeit in the context of Einstein gravity. It will be worthwhile to explore whether the same result applies to other gravity theories as well. This will establish our formalism in a broader framework. Even though there can be a large class of theories which would satisfy the diffeomorphism invariance property and qualify as a gravitational Lagrangian, they can be distinguished by the additional criteria of providing second order field equations under variation. One can show that such Lagrangians are unique and are known as the \LL Lagrangians  \cite{Lovelock:1971yv}. The \LL Lagrangians are polynomials in the Riemann curvature tensor, suitably contracted with the completely antisymmetric determinant tensor. A \LL Lagrangian of order $m$ will involve product of $m$ Riemann curvature tensors. The case $m=1$ corresponds to the Einstein gravity, while $m=2$ is known as the Gauss-Bonnet gravity. Incidentally, the Gauss-Bonnet term appears quiet naturally from the low energy effective action of string theory as well (\cite{Zwiebach:1985uq} and references therein). Thus, so to speak, it has a `quantum' origin. We have already elaborated on the physical process version of the first law in dynamical situations for Einstein gravity and shall explore the corresponding scenario when the Gauss-Bonnet term is present. The action for the Einstein-Gauss-Bonnet gravity in $D$ spacetime dimensions reads,
\begin{align}
\mathcal{A}=\int d^{D}x\sqrt{-g}\frac{1}{16\pi}\Big\{R+\alpha \left(R^{2}-4R_{ab}R^{ab}+R_{abcd}R^{abcd}\right) \Big\}~,
\end{align}
where $\alpha$ is a dimensionful coupling constant with dimension $(\textrm{length})^{2}$. One can write down an expression for the entropy associated with black holes in Einstein-Gauss-Bonnet gravity as \cite{Jacobson:1993xs},
\begin{align}
S_{\rm EGB}\equiv  \frac{1}{4}\int d^{D-2}x\sqrt{q}~\left(1+\rho\right);\qquad \rho=2\alpha~^{(D-2)}R~,
\end{align}
where $~^{(D-2)}R$ stands for the intrinsic Ricci scalar of the $(D-2)$ dimensional cross section of the horizon. Note that when $\alpha =0$, $\rho$ vanishes and one recovers the $(\textrm{area}/4)$ law for \gr. One can easily compute the change in entropy due to the dynamical evolution of the event horizon triggered by in falling matter. This can be computed by evaluating the  change in the entropy along the null generator $k^{a}$ parametrized by affine parameter $\lambda$ and leads to the expression,
\begin{align}
\Delta S_{\rm EGB}=\frac{1}{4}\int d^{D-2}x\int _{\lambda _{1}}^{\lambda _{2}}d\lambda~\sqrt{q}\left[\left(1+\rho\right) \theta _{(k)}+\frac{d\rho}{d\lambda}\right]\equiv \frac{1}{4}\int d^{D-2}x\int _{\lambda _{1}}^{\lambda _{2}}d\lambda~\sqrt{q}~ \vartheta_{k}~,
\end{align}
where $\vartheta_{k}$ is the analog of the expansion $\theta_{k}$ in Einstein-Gauss-Bonnet gravity. One can apply the integration by parts method and hence separate out a total divergence from the above integral. This essentially results in,
\begin{align}\label{BH_Dyn_04a}
\Delta S_{\rm EGB}=\Delta \Big\{\frac{1}{4}\int d^{D-2}x~\lambda \sqrt{q} ~\vartheta _{k} \Big\}-\frac{1}{4}\int d^{D-2}x\int _{\lambda _{1}}^{\lambda _{2}}d\lambda~\lambda \sqrt{q}\,\frac{d\vartheta_{k}}{d\lambda}~.
\end{align}
In order to arrive at the above expression we have neglected all the quadratic terms in $\theta _{(k)}$, as we are working under quasi-stationary approximation. Among the terms in the boundary, the $d\rho/d\lambda$ term essentially depends on how $~^{(D-2)}R$ changes along the null generators. Since $~^{(D-2)}R$ depends on the metric $q_{ab}$ projected on to the $(D-2)$ dimensional cross section of the horizon it follows that change of $~^{(D-2)}R$ is essentially related to $dq^{ab}/d\lambda$. This is nothing but the extrinsic curvature of the surface generated by the space of null generators. Thus one obtains,   
\begin{align}\label{BH_Dyn_03a}
\frac{d}{d\lambda}~^{(D-2)}R=-2~^{(D-2)}R_{ab}\left(\sigma ^{ab}_{(k)}+\frac{\theta _{(k)}}{D-2}q^{ab} \right)~,
\end{align}
where $\sigma ^{ab}_{(k)}$ is the shear tensor associated with the null generator $k^{a}$. Regarding the bulk term, which corresponds to $d\vartheta_{k}/d\lambda$, it is possible to simplify it considerably by keeping in mind the approximations under which we are working. In particular one can show that this term is nothing but $-8\pi T_{ab}k^{a}k^{b}$ 
\cite{Kolekar:2012tq}, alike the corresponding scenario in \gr. Finally, we arrive at the following expression for the change in entropy for the Einstein-Gauss-Bonnet gravity,
\begin{align}
\Delta S_{\rm EGB}=\frac{1}{4}\Delta \Big\{\int d^{D-2}x\sqrt{q}~\lambda ~\vartheta_{k} \Big\}
+2\pi\int d^{D-2}x\int _{\lambda _{1}}^{\lambda _{2}}d\lambda~\lambda \sqrt{q}\,T_{ab}k^{a}k^{b}~.
\end{align}
However, the above result is in affine parametrization and therefore it is necessary to make a transition to a non-affine parametrization following the routes described in the earlier section. Further since $\lambda$ appears multiplied with the expansion and we are working under linear approximation, one can take it to be the one corresponding to the Killing vector of the stationary background. When these are taken into account, the change in entropy due to the evolution of the horizon along the non-affinely parametrized null generators can be written as, 
\begin{align}\label{BH_Dyn_04c}
\frac{\kappa}{2\pi }  \Delta S_{\rm EGB}&=\Delta \Big\{\frac{1}{8\pi }\int d^{D-2}x\,\sqrt{q} ~\vartheta \Big\}
+ \int_{\cal H} T_{ab}\xi^{a}d\Sigma^{b}~.
\end{align}
Note that if the end points correspond to past bifurcation surface and future stationary slice, then the boundary contributions will vanish as in case of \gr\ and one ends up with integrated version of the physical process first law for Einstein-Gauss-Bonnet gravity \cite{Chatterjee:2011wj}. In situations where we compute the entropy change between arbitrary cross sections, the boundary terms come into play and it is needed to interpret them in order for \ref{BH_Dyn_04c} to have a proper thermodynamic significance. Before venturing into such an analysis, we note that the term $\vartheta$ also represent the analog of expansion in this case, measuring the change in entropy per unit cross section. For Einstein-Gauss Bonnet theory, the entropy has been shown to be increasing at any arbitrary slice for linearized perturbation with positive energy matter. Therefore, as in the case of \gr, $\vartheta$ is also positive semi-definite at least for linearized perturbation of the horizon. Next, taking a cue from the situation with \EH action we try to compare the boundary term with the energy of the horizon fluid in membrane paradigm, but now for Einstein-Gauss-Bonnet gravity. The membrane stress energy tensor $t_{ab}$ for Einstein-Gauss-Bonnet gravity has been obtained in \cite{Jacobson:2011dz} for perturbations over a spherically symmetric black hole horizon, and the corresponding energy density $\Sigma =t_{ab}u^{a}u^{b}$ associated with an evolving horizon in the membrane paradigm is,
\begin{align}
\Sigma _{\rm EGB}= -\frac{\theta}{8\pi}\left[1+2\alpha \frac{D-4}{D-2}~^{(D-2)}R\right]~.
\end{align}
When $\alpha \rightarrow 0$, it follows that $\Sigma =-\theta/8\pi$, reproducing the \gr\ result for membrane energy density. We then evaluate $\vartheta$ for the case of linearized perturbations around a spherically symmetric horizon leading to: 
\begin{align}
\vartheta = \Sigma _{\rm EGB}= -\frac{\theta}{8\pi}\left[1+2\alpha \frac{D-4}{D-2}~^{(D-2)}R\right]~.
\end{align}
This allows us to finally write down a first law for entropy change between two arbitrary cross section of the dynamical horizon in Einstein-Gauss Bonnet theory as,
\begin{align}
\frac{\kappa}{2\pi} \Delta S=\Delta E +  \int_{\cal H} T_{ab}\,\xi^{a}d\Sigma^{b} ;\qquad E=-\int d^{D-2}x\sqrt{q}\,\Sigma _{\rm EGB}~.
\end{align}
This again coincides with the interpretation put forward in the context of \gr. Thus in dynamical situations the matter flow into a horizon in Einstein-Gauss-Bonnet gravity changes the entropy and the energy associated with the horizon. This provides a concrete realization of the first law and illustrates the inter-relationship of black hole thermodynamics and membrane paradigm both in the context of Einstein as well as Einstein-Gauss-Bonnet gravity. This prompts us to see whether an identical interpretation can be put forward for general Lovelock theories. This is what we will do next. 
\section{Generalization to the full \LL Lagrangian}\label{Sec_04}

Having described a dynamical version of the physical process first law for entropy change between arbitrary cross sections, in the context of Einstein-Gauss-Bonnet gravity, we would like to be more ambitious and try to see whether a similar result persists for the full \LL Lagrangian. As already mentioned earlier, the full \LL Lagrangian is a sum over various polynomials of the Riemann curvature and it can be presented as,
\begin{align}\label{BH_Dyn_LL}
L_{\rm LL}=\frac{1}{16\pi}\sum _{m=1}c_{m}\delta ^{a_{1}b_{1}\cdots a_{m}b_{m}}_{c_{1}d_{1}\cdots c_{m}d_{m}} R^{c_{1}d_{1}}_{a_{1}b_{1}}\cdots R^{c_{m}d_{m}}_{a_{m}b_{m}}\equiv \frac{1}{16\pi}\sum _{m=1}c_{m}L_{(m)}~.
\end{align}
Here $c_{m}$'s are arbitrary dimensionful coupling coefficients and $\delta ^{ab\cdots}_{cd\cdots}$ represent the completely antisymmetric determinant tensor. The first term in the above Lagrangian corresponds to the Ricci scalar, while the second one stands for the Gauss-Bonnet invariant. The \LL\ class of gravity theories have been extensively analyzed in the context of black hole mechanics \cite{Padmanabhan:2013xyr, Chakraborty:2015wma}. As a warm up, we will first illustrate the first law for a single term in the above Lagrangian, known as pure Lovelock terms and shall subsequently generalize to the full \LL theory. 
\subsection{The case of pure \LL gravity}

In this section we will exclusively work with the $m$th order \LL gravity which is a product of $m$ Riemann tensors suitably contracted with the determinant tensor, given by $L_{(m)}$ as in \ref{BH_Dyn_05}. For the $m$th order \LL Lagrangian, the corresponding entropy associated with a horizon corresponds to,
\begin{align}\label{BH_Dyn_06}
S^{(m)}_{\rm LL}\equiv \frac{1}{4}\int d^{D-2}x\sqrt{q}~\rho~;\qquad \rho=m~^{(D-2)}L_{(m-1)}~.
\end{align}
Thus the entropy of a $(D-2)$ dimensional cross section of a horizon in $m$th order Lovelock theory is proportional to the $(m-1)$th \LL Lagrangian in $(D-2)$ dimensions 
\cite{Jacobson:1993xs}. One can immediately compute the associated change in the entropy of the horizon as some matter falls into the same. This can be obtained by computing how the entropy changes along the null generator $k^{a}$ associated with the horizon. Thus the change in entropy, to leading order, would correspond to \ref{BH_Dyn_04a}, where $\rho$ is now given by \ref{BH_Dyn_06}. The term $d\rho/d\lambda$ in this context depends on how $~^{(D-2)}L_{(m-1)}$ changes along the null generator $k^{a}$. This again corresponds to $~^{(D-2)}\mathcal{R}_{ab}^{(m-1)}(dq^{ab}/d\lambda)$. Here $~^{(D-2)}\mathcal{R}_{ab}^{(m-1)}$ is the analog of Ricci scalar for the $(m-1)$th \LL Lagrangian in $(D-2)$ spacetime dimensions. Note that for Gauss-Bonnet gravity $m=2$ and one recovers the standard Ricci scalar which is exactly what we have shown in the previous section. Again 
using the fact that the change of $q_{ab}$ along $\lambda$ is related to the extrinsic curvature associated with the horizon surface, it follows that,
\begin{align}
\frac{d}{d\lambda}~^{(D-2)}L_{(m-1)}=-2~^{(D-2)}\mathcal{R}_{ab}^{(m-1)}\left(\sigma ^{ab}_{(k)}+\frac{\theta _{(k)}}{D-2}q^{ab} \right)~.
\end{align}
Since, ultimately we will consider perturbations of a spherically symmetric black hole, the shear part does not contribute in the leading order and we obtain $(d/d\lambda)\{~^{(D-2)}L_{m-1}\}$ as $\{-2(m-1)/(D-2)\}\theta _{(k)}~^{(D-2)}L_{m-1}$. Note that for the case when $m=2$, the result for Gauss-Bonnet gravity is retrieved. Further, the derivation of the second law for linearized perturbations in \LL gravity \cite{Kolekar:2011gw} shows that the bulk term $d\vartheta _{k}/d\lambda$ appearing in the entropy change as in \ref{BH_Dyn_04a} is  again equal to $-8\pi T_{ab}k^{a}k^{b}$. Using these results one finally arrives at the change of entropy between two arbitrary cross sections of the horizon as,
\begin{align}\label{BH_Dyn_07}
\Delta S^{(m)}_{\rm LL}=\Delta \left\{ \frac{m}{4} \int d^{D-2}x~ \lambda \sqrt{q}\,\frac{D-2m}{D-2}\theta _{(k)}~^{(D-2)}L_{m-1}\right\}
+2\pi\int d^{D-2}x\int _{\lambda _{1}}^{\lambda _{2}}d\lambda~\lambda \sqrt{q}\,T_{ab}k^{a}k^{b}
\end{align}
The membrane paradigm for black holes in \LL gravity has been studied for linearized perturbations of spherically symmetric black holes \cite{Kolekar:2011gg} and the membrane stress tensor is of the form of a viscous fluid as in case of \gr. The  energy density associated with membrane stress-energy tensor for $m$th order Lovelock Lagrangian then reads,
\begin{align}\label{BH_Dyn_08a}
\Sigma ^{(m)}_{\rm LL}=-\frac{m}{8\pi}\frac{D-2m}{D-2}\theta~^{(D-2)}L_{m-1}
\end{align}
Thus one can rewrite \ref{BH_Dyn_07} using non-affine parametrization and the energy density of the membrane fluid in Lovelock theories to finally arrive at,
\begin{align}
\frac{\kappa}{2 \pi} \Delta S_{\rm LL}^{(m)}=\Delta E + \int_{\cal H} T_{ab}\,\xi^{a}d\Sigma^{b} ;\qquad E=-\int d^{D-2}x\sqrt{q}\,\Sigma ^{(m)}_{\rm LL}
\end{align}
Thus even in the context of pure \LL theories one can arrive at a physical process version of first law in dynamical situations, leading to an identical description for the change in entropy between arbitrary non-equilibrium states as in previous scenarios. We will now collect all these results and generalize this physical process version of first law to the full \LL gravity.
\subsection{Physical Process First law and the \LL gravity}

In this section we will consider the full \LL Lagrangian given by \ref{BH_Dyn_05}.  One can compute the entropy associated with a horizon in the context of full \LL Lagrangian, which corresponds to, 
\begin{align}\label{BH_Dyn_08}
S_{\rm LL}=\sum _{m=1}S_{\rm LL}^{(m)}=\frac{1}{4}\int d^{D-2}x\sqrt{q}\left(1+\rho\right);\qquad \rho \equiv -2P^{abcd}k_{a}\ell _{b}k_{c}\ell _{d}-1=\sum _{m=2}c_{m}m\{~^{(D-2)}L_{m-1}\}
\end{align}
where the notations and constants have the usual meaning. For the \EH Lagrangian $\rho=0$, while for other Lagrangians it is non-trivial. For example, in Einstein-Gauss-Bonnet gravity one would have obtained $\rho=2\alpha~^{(D-2)}R$, where $\alpha$ is the coupling parameter associated with the Gauss-Bonnet Lagrangian, which we have explicitly used in \ref{Sec_03}. Taking a cue from our earlier analysis, the change in entropy would correspond to $\left(1+\rho\right) \theta _{(k)}+d\rho/d\lambda$. Again one can separate out a total derivative term and rewrite the change of entropy as a difference between two boundary terms and a bulk term. The form is structurally identical with \ref{BH_Dyn_04a}, with $\rho$ being given by \ref{BH_Dyn_08}. Following \cite{Kolekar:2011gg}, it turns out that the $d\vartheta _{k}/d\lambda$ term in \ref{BH_Dyn_04a} will lead to $-8\pi T_{ab}k^{a}k^{b}$, while the boundary term can be obtained following the pure Lovelock analysis (see e.g., \ref{BH_Dyn_08a}) such that,
\begin{align}
\theta _{k}\left(1+\rho\right)+\frac{d\rho}{d\lambda}=\theta _{k}\left[1+\sum _{m=2}mc_{m} \frac{D-2m}{D-2}~^{(D-2)}L_{m-1} \right]
=-\frac{8\pi}{\kappa \lambda} \Sigma _{\rm LL}
\end{align}
where $\Sigma _{\rm LL}$ is the energy density associated with the horizon fluid in the membrane paradigm. Transforming to non-affine parametrization, one can finally write down the following structure of the dynamical version of physical process first law in full \LL gravity as,
\begin{align}
\frac{\kappa}{2 \pi} \Delta S _{\rm LL}=\Delta E + \int_{\cal H} T_{ab}\,\xi^{a}d\Sigma^{b} \qquad
E=-\int d^{D-2}x\sqrt{q}~\Sigma _{\rm LL}
\end{align}
Thus we arrived at an identical interpretation for the change in entropy associated with a horizon due to matter flowing into it. It further depicts that the energy going into the horizon in the form of matter stress energy tensor results into a change in black hole entropy and also modifies the energy density associated with horizon fluid in membrane paradigm. Interestingly, if the end points of this physical process corresponds to past bifurcation surface and future stationary cross section then the energy term identically vanishes at these two boundaries and one arrived at the standard integrated physical process first law.   
\section{Conclusion}
Generically, realistic black holes are supposed to be dynamic due to presence of flux across the horizon. It can however be stationary for finite intervals of time when there is no flux. Hence the notion of Black holes as depicted by global stationary solutions of some equations of motion describing gravity, does not exist in reality. Consequently  the assumption of existence of a regular bifurcation surface associated with every non-extremal black hole is only an idealization. Therefore there is an immediate difficulty in defining the entropy of a black hole in such scenarios. As has been discussed this is due to the ambiguities in the assigning of Wald's entropy to the cross-sections of such dynamical black holes. We have therefore discussed the ambiguities in the definition of Wald's entropy and the extent to which it affects the PPFL. In particular we show that to linear orders in perturbation the ambiguities in the Wald entropy do not affect the PPFL as long as the integration range is from the initial bifurcation surface to the final stationary horizon cross-section. However, the second order variation of entropy turns out to be affected by ambiguities.

Given a non degenerate Killing horizon $\mathcal K$, there are theorems that state that one can always extend the spacetime such that $\mathcal K$, in the extended spacetime is a bifurcate Killing horizon \cite{Racz:1995nh}, hence establishing the fact that on a larger manifold the assumption of the presence of a bifurcation surface is viable. Therefore in the dynamical scenario, provided the black hole was stationary initially, one may assume the presence of a bifurcation surface. On the other hand, it also seems reasonable that a version of the physical process first law for dynamical black holes must hold even without the assumption of the existence of a bifurcation surface. It is precisely this question that we answer in the affirmative, within the dynamical event horizon framework. This is a well motivated question considering the fact that in certain quasi-local definitions of evolving black holes viz. dynamical horizons which do not assume a bifurcation surface, an integrated form of the first law can be realized \cite{Ashtekar:2002ag, Ashtekar:2003hk} \footnote{Certain differential forms of the first law also hold under a slow evolution condition \cite{Booth:2003ji}}. 

The main outcome of the second part of the work is being able to write the first law for entropy change between two non equilibrium cross sections at the expense of introduction of the extra term $\Delta E$ which has an interpretation as the change of the energy of the horizon fluid in the context of the black hole membrane paradigm. Note that, we can not assert the sign of the membrane energy unless we demand that the expansion is positive at every cross section. This is indeed true because of the Hawking's area theorem. But the area theorem requires the assumption of cosmic censorship. Therefore, we indirectly require the stability of the black hole under perturbation even if we are interested only in the change of entropy between two intermediate cross sections. 

Consider the situation in \gr, where the focusing equation along with null energy condition forces the expansion to decrease as the matter flux  perturbs the horizon. So, the expansion remains positive but decreases to zero in the future and the matter flux is used up for two processes. First to increase the entropy of the non equilibrium cross sections of the horizon, and secondly to decrease the expansion and increase the membrane energy from negative to zero. It is indeed interesting that all these can be easily extended to higher curvature theories like \LL gravity. This seems to indicate that the thermodynamic nature of horizon is an universal property independent of the detailed dynamics of gravity.

There are several possible extensions of our result. We hope that the formalism may be useful to derive a physical process first law for extremal black holes. Also, except \gr, the membrane paradigm is developed only for perturbations about a spherically symmetric horizon.  It would be interesting if more general cases can be compared. Another important extension could be to go beyond the linear order of perturbation and understand the effect of dissipative terms.
\section*{Acknowledgements}

Research of SC is funded by the INSPIRE Faculty Fellowship (Reg. No. DST/INSPIRE/04/2018/000893)
from Department of Science and Technology, Government of India. AG is supported by SERB, government of India through the NPDF grant (PDF/2017/000533). SS is supported by the Department of Science and Technology, Government of India under the SERB Fast Track Scheme for Young Scientists (YSS/2015/001346). SC thanks IIT Gandhinagar for hospitality, where part of this work was carried out. 

\end{document}